\documentclass[reprint,12pt]{elsarticle}

\usepackage{amssymb, amsmath,bm}
\usepackage{graphicx}
\usepackage{graphics}
\usepackage{latexsym}
\usepackage{booktabs}
\journal{Physics Letters B}

\begin{document}

\begin{frontmatter}



\title{Consistent large-scale shell-model analysis of the two-neutrino $\beta\beta$ and 
single $\beta$ branchings in $^{48}\rm Ca$ and $^{96}\rm Zr$}


\author{Joel Kostensalo and Jouni Suhonen}

\address{{ \small \it University of Jyvaskyla,
Department of Physics, P. O. Box 35, FI-40014, Finland.}}

\begin{abstract}

Two-neutrino double-beta-decay matrix elements $M_{2\nu}$ and single beta-decay branching 
ratios were calculated for $^{48}$Ca and $^{96}$Zr in the interacting nuclear shell model
using large single-particle valence spaces with well-tested two-body Hamiltonians. 
For $^{48}$Ca the matrix element $M_{2\nu}=0.0511$ is obtained, which is 5.5\% smaller 
than the previously reported value of 0.0539. For $^{96}$Zr this work reports the first 
large-scale shell-model calculation of the nuclear matrix element, yielding a value 
$M_{2\nu}=0.0747$ with extreme single-state dominance. If the scenario where the first $1^+$ state
in $^{96}$Nb turns out to be correct, the matrix element is increased to 0.0854.
These matrix elements, combined with the available $\beta\beta$-decay 
half-life data, yield effective values of the weak axial coupling which in turn are used
to produce in a consistent way the $\beta$-decay branching ratios of $(7.5\pm2.8)$ \% 
for $^{48}$Ca and $(18.4\pm0.09)$ \% for $^{96}$Zr. These are larger than obtained in previous 
studies, implying that the detection of the $\beta$-decay branches could be possible
in dedicated experiments sometime in the (near) future.

\end{abstract}

\begin{keyword}

double-beta decay \sep axial-vector coupling \sep 48Ca \sep 
96Zr \sep shell model \sep matrix elements



\end{keyword}

\end{frontmatter}



%% INTRO
%\section{\label{sec:intro}Introduction} 

The nuclei $^{48}$Ca and $^{96}$Zr share an interesting feature, the two being the 
only known nuclei where single $\beta$-decay transitions compete with the dominant two-neutrino 
double beta ($2\nu\beta\beta$) decay \cite{Ejiri2019}. 
This exceptional situation is due to the large 
angular-momentum difference ($\Delta J=4,5,6$) between the initial and final states 
of $\beta$ decays, as well as the relatively small decay energies ($Q$ values). Both 
theoretical and experimental studies have been carried out regarding decays of both nuclei 
\cite{Poves1995,Suhonen1993,Balysh1996,Bakalyarov2002,Haaranen2014,Heiskanen2007,Alanssari2016}. 
The two-neutrino-emitting modes are dominated by the ground-state-to-ground-state transitions 
with resent half-life estimates of $6.4^{+1.4}_{-1.1}$ for $^{48}$Ca \cite{Arnold2016} and 
$2.35\pm 0.21$ for $^{96}$Zr \cite{Argyriades2010}. These two nuclei are favorable for 
experimental double-beta-decay studies due to their large $Q$ values: $^{48}$Ca having 
the largest known double-$\beta$ $Q$ value $Q_{\beta\beta}(^{48}\rm Ca)=4269.08(8)$ keV, 
and $^{96}$Zr having the third largest value $Q_{\beta\beta}(^{96}\rm Zr)=3356.03(7)$ keV, 
with only $^{150}$Nd between them \cite{nndc}.

The single-$\beta$ channels have not yet been observed but lower limits for the half-lives 
stand at $1.1\times10^{20}$ yr for $^{48}$Ca \cite{Bakalyarov2002} and $2.6\times 10^{19}$ yr 
for $^{96}$Zr \cite{Barabash1996}. Were these observed, they would provide valuable 
information about the validity of current nuclear models, which could be used to improve 
the accuracy of calculations of the matrix elements of neutrinoless $\beta\beta$ decay. 

In this Letter we revisit the previous theoretical studies, giving an updated estimate 
for the $2\nu\beta\beta$-decay matrix element for $^{48}$Ca and for the first 
time a shell-model estimate for the $^{96}$Zr $2\nu\beta\beta$-decay matrix element. 
Using this information we present improved estimates for the $\beta$-decay branching 
ratios. This knowledge can in the future be used to design optimal experiments for 
the detection of the $\beta$-decay branches.

%This article is organized as follows. In Sec. \ref{sec:theory} we give an overview of 
%the theory of $\beta$ and $\beta\beta$ decays, in Sec. \ref{sec:calc} we give the 
%necessary details of our calculations, in Sec. \ref{sec:results} we present our 
%results, and in Sec. \ref{sec:conc} we draw conclusions. 

%THEORY
%\section{\label{sec:theory}Theoretical Background} 

%In this section we review the basic theory behind calculating single and 
%double $\beta$-decay half-lives. 
The theory of $\beta$ decay, including the 
forbidden transitions considered here, is extensively treated in the work of 
Behrens and B\"uhring \cite{Behrens1982}. A streamlined presentation of the theory, 
including all the technical details of how the calculations were carried out 
also in the present work, can be found from \cite{Haaranen2017}. The basic 
theory behind the $2\nu\beta\beta$ decay can be found in much more detail for example 
from \cite{Suhonen1998}.

For $\beta$ decay the probability of the electron being 
emitted with kinetic energy between $W_e$ and $W_e+dW_e$ is
\begin{align}
\notag
P(W_e)dW_e =& \frac{G_{\rm F}}{(\hbar c)^6}\frac{1}{2\pi^3\hbar}C(W_e) \\
& \times p_ecW_e(W_0-W_e)^2F_0(Z,W_e)dW_e,
\label{eq:emisprob}
\end{align}
where $p_e$ is the momentum of the electron, $Z$ is the proton number of the final-state nucleus, 
$F_0(Z,W_e)$ is the so-called Fermi function, and $W_0$ is the end-point energy of 
the $\beta$ spectrum. The nuclear-structure information is encoded as form factors in 
the shape factor $C(w_e)$. In the impulse approximation, where we assume that the 
decaying nucleon does not interact with the other $A-1$ nucleons at the moment of decay, 
these form factors map to nuclear matrix elements (NMEs), which can in turn be calculated 
using a many-body framework, such as the interacting nuclear shell model. The axial-vector 
coupling $g_{\rm A}$ and the vector coupling $g_{\rm V}$, which enter 
the theory of $\beta$ decay when the vector and axial-vector hadronic currents become 
renormalized at the nucleon level, appear as multipliers of the various axial-vector and 
vector matrix elements respectively.

The half-life of $\beta$ decay can be written as
\begin{equation}
t_{1/2} = \frac{\kappa}{\tilde{C}},
\label{eq:half-life}
\end{equation}
where $\tilde{C}$ is the integrated shape function and the constant $\kappa$ has 
the value \cite{Hardy1990}
\begin{equation}
\kappa = \frac{2\pi^3\hbar^7\mathrm{ln \  2}}{m_e^5c^4(G_{\rm F}
\cos \theta_{\rm C})^2}= 6147 \ \mathrm{s},
\label{eq:kappa}
\end{equation}
$\theta_{\rm C}$ being the Cabibbo angle.
To simplify the formalism it is traditional to introduce unitless kinematic 
quantities $w_e=W_e/m_ec^2$, $w_0 = W_0/m_ec^2$ and 
$p=p_ec/(m_ec^2)= \sqrt{w_e^2 -1}$, and so the integrated 
shape function can then be expressed as
\begin{equation}
\tilde{C} = \int^{w_0}_1 C(w_e)pw_e(w_0-w_e)^2F_0(Z,w_e)dw_e.
\label{eq:ctilde}
\end{equation} 
The shape factor $C(w_e)$ of Eq.~(\ref{eq:ctilde}) contains complicated combinations of
both (universal) kinematic factors and NMEs. As in the previous studies regarding 
forbidden $\beta$ decays \cite{Haaranen2017,Haaranen2016,Kostensalo2017} we take into 
account the next-to-leading-order terms of the shape factor as well as screening and 
radiative effects.  

For the $2\nu\beta\beta$ decay the half-life expression is analogous to that of $\beta$ decay 
in Eq. (\ref{eq:half-life}) and can be written as \cite{Suhonen1998}
\begin{equation}
t^{(2\nu)}_{1/2} = \frac{1}{G^{(2\nu)}g_{\rm A}^4|M_{2\nu}|^2},
\label{eq:2vbbhl}
\end{equation}
where $G^{(2\nu)}$ is the phase-space integral (the expression for this is given in, 
e.g. \cite{Suhonen1998}) and $M_{2\nu}$ is the matrix element given for 
$\beta^-\beta^-$ decay by
\begin{equation}
M_{2\nu} = \sum_m \frac{(0_{\rm g.s.}^{(f)}|| \sigma\tau^- || 1^+_m)
( 1^+_m|| \sigma\tau^- || 0_{\rm g.s.}^{(i)} )}{[\frac{1}{2}Q_{\beta\beta} + E(1^+_m)-M_i]/m_e +1},
\end{equation}
where $m_e$ is the electron rest mass, $E(1^+_m)-M_i$ is the energy difference between 
the $m$th intermediate 1$^+$ state and the ground state of the initial nucleus, and 
$Q_{\beta\beta}$ is the energy released in the decay (i.e. $Q$ value).

% CALCULATIONS
%\section{\label{sec:calc} Shell-model calculations}

The nuclear-structure calculations were done using the interacting shell model with 
the computer code NuShellX@MSU \cite{nushellx}. Following the earlier shell-model studies 
regarding the half-lives of the transitions 
$^{48}\rm Ca(0^+) \rightarrow \, ^{48}\rm Sc(4^+,5^+,6^+)$ \cite{Haaranen2014} and the 
$2\nu\beta\beta$-decay channel \cite{Horoi2007}, the full $fp$ model space with the 
interaction GXPF1A \cite{Honma2004,Honma2005} was used.

For the decay of $^{96}$Zr a model space including the 
proton orbitals $0f_{5/2}$, $1p_{3/2}$, $1p_{1/2}$ and $0g_{9/2}$ and the neutron 
orbitals $0g_{7/2}$ $1d_{5/2}$, $1d_{3/2}$ and $0s_{1/2}$ were used together with the 
interaction glekpn \cite{Mach1990}. In the previous shell-model study \cite{Alanssari2016} 
the calculations were done in the much smaller proton $0g_{9/2}$--$1p_{1/2}$ and 
neutron $1d_{5/2}$--$2s_{1/2}$ model space with the Gloeckner interaction \cite{Gloeckner1975}. 
While the exclusion of a large number of important orbitals can affect the accuracy of 
the computed half-lives of the various $\beta$-decay branches, the problem is even 
more severe for the ground-state-to-ground-state $2\nu\beta\beta$ decay, which is 
strictly forbidden in such a limited model space. In the present study this transition can 
proceed by simultaneous Gamow-Teller transitions between the proton $0g_{9/2}$ 
and neutron $0g_{7/2}$ orbitals.

Since the computational burden for description of these decays is manageable for modern 
computers, we included all the intermediate $1^+$ states of $2\nu\beta\beta$ decay in $^{48}$Sc 
and $^{96}$Nb. This is an improvement over the previous calculation regarding 
the matrix element of $^{48}$Ca \cite{Horoi2007}, where only 250 intermediate states 
were used. For $^{48}$Sc our extended calculation includes 9470 $1^+$ states and 
excitation energies up to 60 MeV, while 
for $^{96}$Nb we have 5894 $1^+$ states reaching energies of roughly 18 MeV. Since the 
exact energies of the intermediate states play an important role in the determination 
of the $2\nu\beta\beta$ NMEs, the excitation energies of the $1^+$ states 
in $^{48}$Sc were shifted such that the lowest-lying state is at the experimental energy of 
2200 keV \cite{nndc}. For $^{96}$Nb no $1^+$ states are known experimentally, so that
the shell-model excitation energies were used. However, the paper by Thies \textit{et al.} \cite{Thies2012} 
suggests that the state at 694.6 keV is the lowest $1^+$ state. The branching ratio calculations were repeated also 
for this scenario.

The phase-space integrals are taken from the work of Neacsu and Horoi \cite{Neacsu2016}. 
The $Q$ values are taken from \cite{nndc} and are $Q_{\beta\beta}(^{48}\rm Ca)=4269.08(8)$ keV, 
$Q_{\beta}(^{48}\rm Ca)=279(5)$ keV, $Q_{\beta\beta}(^{96}\rm Zr)=3356.03(7)$ keV, 
and $Q_{\beta}(^{96}\rm Zr)=163.97(10)$ keV.

% RESULTS
%\section{\label{sec:results} Results} 

In the following we will first report on the computed results for the $2\nu\beta\beta$ NMEs of 
$^{48}$Ca and $^{96}$Zr and then extract effective values $g_{\rm A}^{\rm eff}$ of the weak axial 
coupling based on comparisons with the measured $2\nu\beta\beta$ half-lives. 
These $g_{\rm A}^{\rm eff}$ are then, in turn, used to predict the $\beta$-decay branching 
ratios for transitions to the lowest $4^+$, $5^+$, and $6^+$ states of $^{48}$Sc and $^{96}$Nb.
This we consider to be a consistent approach since the $2\nu\beta\beta$ and $\beta$ decays
are low-momentum-exchange processes and thus the related axial couplings are expected to
be quenched by a similar amount \cite{Suhonen2017,Suhonen2019}.

%\subsection{\label{subsec:2vbb} Two-neutrino $\beta\beta$ matrix elements} 

\begin{table}%[H]
\centering
  \caption{Shell-model calculated $2\nu\beta\beta$ NMEs and the 
extracted effective value $g_{\rm A}^{\rm eff}$ of the axial-vector coupling.} 
		%\begin{ruledtabular}

\begin{tabular}{ccccc}
		\hline
   Nucleus & $\vert M_{2\nu}\vert$ & $G$ ($10^{-18}$ yr$^{-1}$) \cite{Neacsu2016}&$T_{1/2}^{\beta\beta}$ 
($10^{19}$ yr)&$g_{\rm A}^{\rm eff}$\\
  \hline
$^{48}\rm Ca$ &0.0511&14.805&$6.4^{+1.4}_{-1.1}$ \cite{Arnold2016} &$0.80\pm0.04$\\
$^{96}\rm Zr$ &0.0747&6.420&$2.35\pm0.21$ \cite{Argyriades2010} &$1.04^{+0.03}_{-0.02}$ \\
\hline
\end{tabular}
\label{tbl:NMEs}

	%\end{ruledtabular}
\end{table}

			\begin{figure}[h!]
	\centering	
	\includegraphics[width=0.5\textwidth]{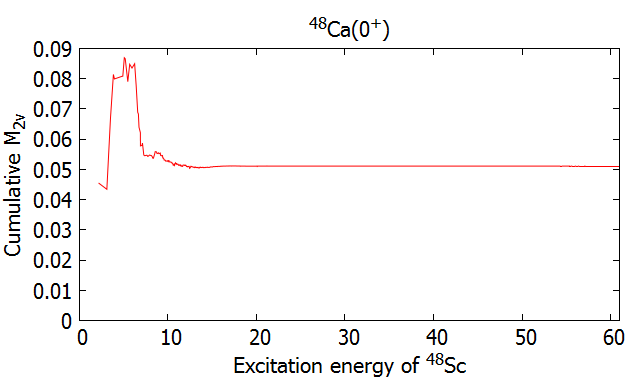}
	\caption{Cumulative $2\nu\beta\beta$ NME
$M_{2\nu}$ for $^{48}$Ca as a function of excitation energy of the intermediate state in $^{48}$Sc.
\label{fig:48cumu}  }
	\end{figure}

			\begin{figure}[h!]
	\centering	
	\includegraphics[width=0.5\textwidth]{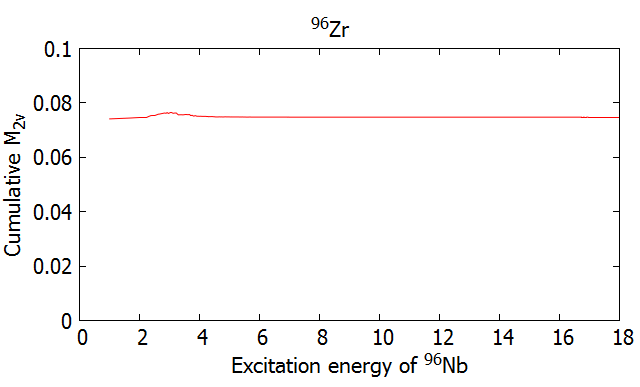}
	\caption{Cumulative $2\nu\beta\beta$ NME 
$M_{2\nu}$ for $^{96}$Zr as a function of excitation energy of the intermediate state in $^{96}$Nb.
\label{fig:96cumu}  }
	\end{figure}

The computed shell-model $2\nu\beta\beta$ NMEs are given in 
table \ref{tbl:NMEs}. For $^{48}$Ca the shell-model calculation gives $|M_{2\nu}|$=0.0511, 
which is 5 \% smaller than the value 0.0539 reported in \cite{Horoi2007}. The 
accumulation of the matrix element is in agreement with the previous results 
(see Fig. \ref{fig:48cumu}). The lowest $1^+$ state is the most important, 
contributing an amount of 0.0454 to the total NME. The next dozen states are mostly 
constructive, adding up to a maximum value of 0.0847 of the NME, beyond which the states 
start to contribute destructively. Our cumulative NME agrees with the previous 
result when 50 intermediate states are used (reaching at about 9.4 MeV). The next 
approximately 150 states (reaching 13.4 MeV) add destructively to the cumulative 
NME bringing the NME to a value 0.0505. The 200th to 460th states (up to 16.9 MeV) 
add constructively beyond which the cumulative matrix element tapers off to the 
final value 0.0511.

In the case of $^{96}$Zr (see Fig. \ref{fig:96cumu}) there is a clear single-state 
dominance (SSD) \cite{Civitarese1998,Civitarese1999}, with the first excited state 
contributing an amount of 0.0747 to the total NME, while the sum of the other contributions 
is zero to three significant digits. This agrees with the measurement  
of Thies \textit{et al.} \cite{Thies2012} where extreme SSD was reported to be found in the 
$2\nu\beta\beta$ NME of $^{96}$Zr in a high-resolution $^{96}$Zr($^3$He,$t$) experiment.
Hence, our calculations confirm the experimental result of \cite{Thies2012}.
The accumulation for $^{96}$Zr is very similar to the $48$Ca 
case with the first 15 states adding constructively to 0.0765, beyond which the rest 
of the states contribute destructively. Beyond the first 100 states (7.5 MeV) the 
contributions are negligible.
 
Solving for $g_{\rm A}$ from equation (\ref{eq:2vbbhl}) and using the experimental 
half-lives from \cite{Arnold2016,Argyriades2010}, phase-space integrals 
from \cite{Neacsu2016}, and the present shell-model NMEs, we get the 
effective $g_{\rm A}$ values $g_{\rm A}^{\rm eff}=0.80\pm0.04$ for $^{48}$Ca and 
$g_{\rm A}^{\rm eff}=1.04^{+0.03}_{-0.02}$ for $^{96}$Zr. These values of $g_{\rm A}^{\rm eff}$ are 
specific for the used model spaces and Hamiltonians. In the work of 
Barea et al. \cite{Barea2013} a relation $g_{\rm A}^{\rm eff}=1.269 \times A^{-0.12}$ between 
the axial-vector coupling and the mass number $A$ was found when analyzing
the values of the $2\nu\beta\beta$ NMEs obtained in earlier calculations using the
interacting shell model. Based on this, we would expect $g_{\rm A}^{\rm eff}=0.80$ for 
$^{48}$Ca and $g_{\rm A}^{\rm eff}=0.73$ for $^{96}$Zr. For calcium the values match perfectly, 
while for zirconium less quenching seems to be needed. However, the $^{96}$Zr value 
seems to be consistent with the recent calculations on $^{130}$Te and 
$^{136}$Xe \cite{Neacsu2015,Horoi2016}, where a value  $g_{\rm A}^{\rm eff}=0.94$ was found, 
since we expect these heavier nuclei to require more $g_{\rm A}$ quenching 
than the lighter $^{96}$Zr. 

The measurement of Thies \textit{et al.} \cite{Thies2012} suggests that the state at 694.6 keV in $^{96}$Nb might be the first $1^+$ state.
If we repeat the calculations with this assumption, the total matrix element for $^{96}$Zr decay is increased to 0.0854, which in turn gives
$g_{\rm A}^{\rm eff}=0.97^{+0.03}_{-0.02}$ for $^{96}$Zr.

%\subsection{\label{subsec:betas} Beta-decay branching ratios} 

The $\beta$ transitions to the $4^+$, $5^+$, and $6^+$ states in $^{48}$Sc and $^{96}$Nb
are 4th-forbidden non-unique, 4th-forbidden unique and 6th-forbidden non-unique, respectively.
A priori, without any calculations, one could predict that the 6th-forbidden non-unique 
$\beta$ transition is much suppressed relative to the other two due to the much higher 
degree of forbiddenness that overwhelms the positive boost coming from the slightly larger 
$Q$ value relative to the other two transitions. With less certainty one could predict that
the NMEs of the two 4th-forbidden $\beta$ transitions are on the same ball park and 
the difference in the $Q$ value is most likely the decisive element in defining the 
branching between the two transitions. In the following we test these hypotheses by the
shell-model calculations of the involved NMEs.
	
			\begin{figure}[h!]
	\centering	
	\includegraphics[width=0.5\textwidth]{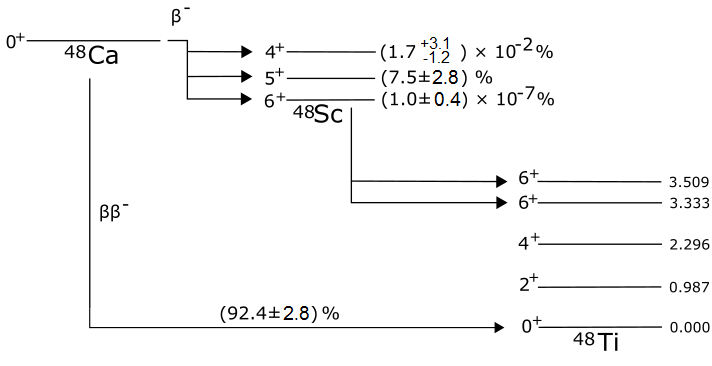}
	\caption{Decay scheme of $^{48}$Ca. Also indicated are our shell-model computed 
$\beta$-decay and $2\nu\beta\beta$-decay branching ratios. 
\label{fig:48scheme}  }
	\end{figure}
		
The $\beta$-decay and $2\nu\beta\beta$-decay branching ratios calculated for $^{48}$Ca 
are indicated in Fig.~\ref{fig:48scheme}. 
As expected, the $2\nu\beta\beta$ branch is clearly dominant with a $92.4\pm2.8$ \% 
branching. The $\beta$-decay branches to the $4^+,5^+,6^+$ states 
are $(1.7^{+3.1}_{-1.2})\times10^{-2}$ \%, $7.5\pm2.8$ \% and $(1.0\pm0.4)\times 10^{-7}$ \%. 
The branching to the $5^+$ state therefore potentially competes with the $2\nu\beta\beta$ branch 
in a significant way, as was pointed out in \cite{Haaranen2014}. The decay to the $6^+$ state is 
greatly hindered by the fact that it is sixth-forbidden. Based on the change in angular 
momentum, we would expect in general the decay to the $4^+$ state be the fastest. However, 
in this case this branch is quite small due to the relatively small $Q$ value. In the work of 
Haaranen et al. \cite{Haaranen2014} the half-lives for the $4^+$ and $6^+$ states were 
reported for $g_{\rm A}=1.0$ and $g_{\rm A}=1.27$. The half-lives are shortened from 
$3.97\times 10^{23}$ yr to $(3.47\pm0.09)\times10^{23}$ yr for the $4^+$ state and 
from $6.39\times10^{28}$ to $5.61\times 10^{28}$ yr, when our $2\nu\beta\beta$-determined
$g_{\rm A}=0.80\pm0.04$ is adopted instead of $g_{\rm A}=1.00$. 
The unique-forbidden $5^+$ branch is  50--60 \% stronger than suggested 
in \cite{Haaranen2014}, since the presently adopted heavier quenching of $g_{\rm A}$ 
affects stronger the $2\nu\beta\beta$ branch. The small differences in the half-lives 
compared to the calculations in \cite{Haaranen2014} are due to the inclusion of the 
next-to-leading-order NMEs and kinematic factors in the present 
study as well as the updated $Q$ value, which is 1 keV larger than used in
the study of \cite{Haaranen2014}. As $g_{\rm A}$ is quenched 
more, the significance of all the $\beta$-decay branches increases. For the unique 
$5^+$ transition the $g_{\rm A}$ dependence of the decay half-life is well known and 
roughly $g_{\rm A}^{-2}$ but for the non-unique transitions this is not the case due to
the more complex structure of the shape factor. The uncertainties related to the branching 
to the $4^+$ state are especially large due to the fact that 
the 5 keV uncertainty makes a large percentage of the 26.65 keV $Q$ value. An accurate 
measurement of the $Q$ value would decrease the uncertainties significantly and 
thus would be desirable. 

			\begin{figure}[h!]
	\centering	
	\includegraphics[width=0.5\textwidth]{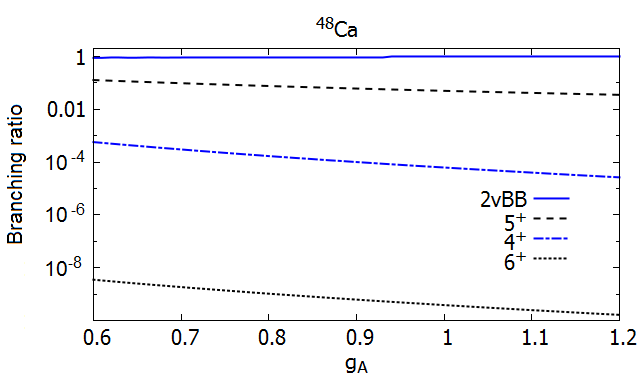}
	\caption{Branching ratios of all the decay branches of $^{48}$Ca as 
functions of $g_{\rm A}$. The solid line represents the $2\nu\beta\beta$-decay 
branching and the dashed and dotted lines the $\beta$-decay branches. The 
$\beta$-decay branches are labeled by the spin-parity of the final state. 
\label{fig:48all}  }
	\end{figure}

			\begin{figure}[h!]
	\centering	
	\includegraphics[width=0.5\textwidth]{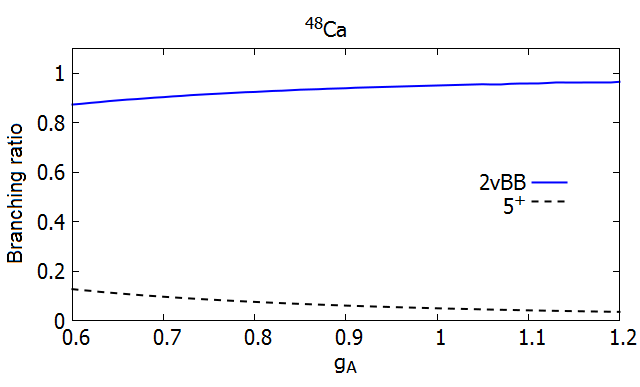}
	\caption{Branching ratios of the two dominant branches of $^{48}$Ca as 
functions of $g_{\rm A}$.
\label{fig:48bb5}  }
	\end{figure}

The $g_{\rm A}$ dependence of all the decay branchings from $^{48}$Ca is studied 
in Fig.~\ref{fig:48all}. As can be seen the dependence on the value of $g_{\rm A}$ is
similar for the $\beta$ branchings and they decrease substantially with 
increasing value of $g_{\rm A}$. At the same time the $2\nu\beta\beta$-decay branching
increases slowly towards one, as can be better seen in Fig.~\ref{fig:48bb5} where only
the $2\nu\beta\beta$ and 4th-forbidden-unique decay branchings are plotted as
functions of $g_{\rm A}$. For reasonable values of $g_{\rm A}$ the $\beta$ branching to 
the $5^+$ state is always below 20 \%.

			\begin{figure}[h!]
	\centering	
	\includegraphics[width=0.5\textwidth]{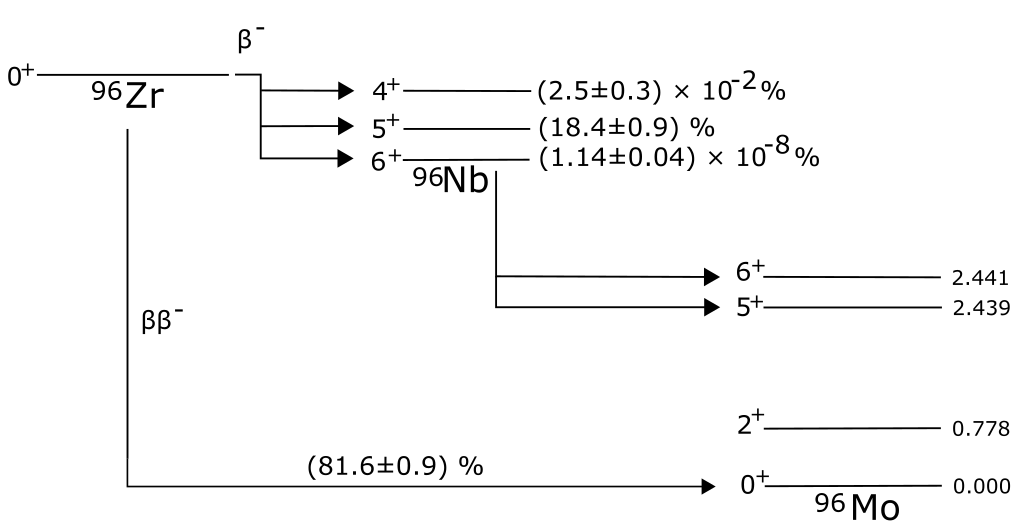}
	\caption{Decay scheme of $^{96}$Zr. Also indicated are our shell-model computed 
$\beta$-decay and $2\nu\beta\beta$-decay branching ratios.
\label{fig:96scheme}  }
	\end{figure}

The computed branching ratios for $^{96}$Zr are presented in Fig. \ref{fig:96scheme}. Like for 
$^{48}$Ca the $2\nu\beta\beta$ branch for $^{96}$Zr is the largest one at 
$(81.6\pm0.9)$ \% but the dominance is not as significant as in the $^{48}$Ca case. 
The branching ratios for the $\beta$-decay transitions are qualitatively 
similar to calcium case with $4^+$ state 
having a branching ratio of $(2.5\pm0.3)\times10^{-2}$ \%, the unique $5^+$ branch 
$(18.4\pm0.9)$ \%, and the ground-state-to-ground-state branch being by far the weakest 
at $(1.14\pm0.04)\times10^{-8}$ \%. Using the computed $\beta$-decay half-lives reported 
in the previous shell-model study \cite{Alanssari2016} of the decay of $^{96}$Zr we can 
extract the corresponding branching ratios of $2.6\times10^{-2}$ \%, $17.6$ \%, and 
$1.21\times10^{-8}$ \% in line with the presently determined branchings. 
The non-unique $\beta$ branches in the present study are slightly smaller than 
those obtained in \cite{Alanssari2016} but the branching to the $5^+$ state is notably 
stronger than expected based on Ref.~\cite{Alanssari2016}. This seems to confirm 
that the $\beta$ decay might be up to 2.3 times faster than predicted by the older 
QRPA calculations in \cite{Heiskanen2007}. 

With the assumption that the first $1^+$ state in $^{96}$Nb is at 694.6 keV, the branching ratios remain
largely unchanged amounting to $(83.3\pm0.8)$ \%, $(3.0\pm0.4)\times10^{-2}$ \%, $(16.7\pm0.9)$ \%, and $(1.19\pm0.15)\times10^{-8}$ \% for the $2\nu\beta\beta$, $4^+$, $5^+$, and $6^+$ decays respectively.

			\begin{figure}[h!]
	\centering	
	\includegraphics[width=0.5\textwidth]{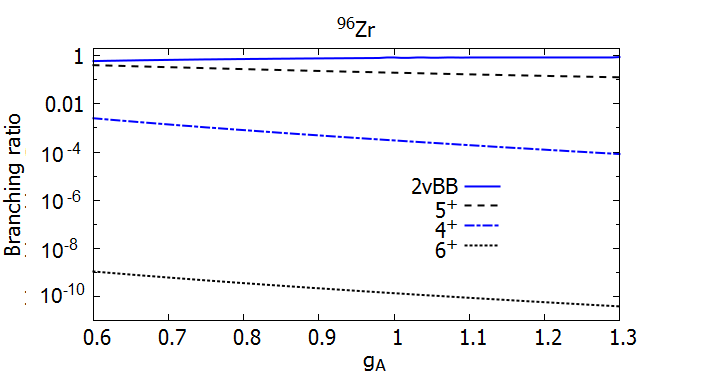}
	\caption{Branching ratios of all the decay branches of $^{96}$Zr as functions 
of $g_{\rm A}$. The solid line represents the $2\nu\beta\beta$-decay branching and the 
dashed and dotted lines the $\beta$-decay branches. The $\beta$-decay branches are labeled 
by the spin-parity of the final state. 
\label{fig:96all}  }
	\end{figure}

			\begin{figure}[h!]
	\centering	
	\includegraphics[width=0.5\textwidth]{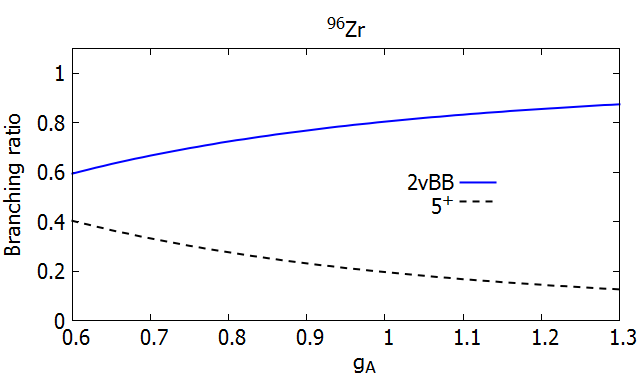}
	\caption{Branching ratios of the two dominant branches of $^{96}$Zr as 
functions of $g_{\rm A}$.
\label{fig:96bb5}  }
	\end{figure}

The dependence of all the $^{96}$Zr decay branchings on $g_{\rm A}$ is studied 
in Fig.~\ref{fig:96all}. As can be seen, a similar $g_{\rm A}$ dependence as in the case 
of $^{48}$Ca is recorded. A closer look at the two leading branchings to the $2\nu\beta\beta$ 
and 4th-forbidden-unique decays, depicted in Fig.~\ref{fig:96bb5}, indicates that their 
$g_{\rm A}$ dependence is much stronger than in the case of $^{48}$Ca. In the $^{96}$Zr
case the $\beta$ branching to the $5^+$ state can reach values up to 40 \% for low values
of $g_{\rm A}$. Such large branchings could be measurable in dedicated experiments sometime
in the future.

% CONCLUSIONS

In this Letter the $2\nu\beta\beta$ matrix elements and single-$\beta$-decay 
branching ratios were calculated for $^{48}$Ca and $^{96}$Zr in the framework of the
interacting nuclear shell model using large single-particle valence spaces and
matching well-tested many-body Hamiltonians. For $^{48}$Ca a $2\nu\beta\beta$ matrix element 
$M_{2\nu}=0.0511$ was obtained, which is 5.5\% smaller than the value of 0.0539 reported 
in the previous calculation of Horoi \textit{et al.} \cite{Horoi2007}. For $^{96}$Zr this was 
the first large-scale shell-model calculation yielding a value of $M_{2\nu}=0.0747$ using the shell model excitation energies and $M_{2\nu}=0.0854$ when the firs $1^+$ state is assumed to be at 694.6 keV in $^{96}$Nb. An extreme single-state dominance was found thus verifying the result of the high-resolution 
$^{96}$Zr($^3$He,$t$) experiment of Thies \textit{et al.} \cite{Thies2012}. Using
these matrix elements, combined with measured values of $2\nu\beta\beta$ half-lives,
effective quenched values of the weak axial coupling $g_{\rm A}$ were extracted to be further
used in the analyses of the $\beta$-decay branchings. In this way a consistent treatment
of both the $2\nu\beta\beta$ decay and the competing $\beta$ decays was achieved.

The $2\nu\beta\beta$-decay and $\beta$-decay branching ratios were studied for 
their $g_{\rm A}$ dependence and the total branchings to the $\beta$ channels were determined
to be $(7.5\pm2.8)$ \% for $^{48}$Ca and $(18.4\pm0.09)$ \% for $^{96}$Zr using the mentioned
consistent effective values of $g_{\rm A}$. These branchings are in both cases larger 
than predicted in previous studies and could be large enough to be detected in underground
experiments in the near future. 
The bulk of the uncertainty related to the $^{48}$Ca branching ratios is due 
to the imprecise knowledge of the $Q$ values. Therefore, a precise measurement of 
the ground-state-to-ground-state $Q$ value for this case would be desirable.   

%\section*{Acknowledgments}\label{thanks}
This work has been partially supported by the Academy of Finland under the Academy 
project no. 318043. J. K. acknowledges the financial support from Jenny and Antti
Wihuri Foundation. 

%\nocite{*}

%\bibliography{BBlit}% Produces the bibliography via BibTeX.

\end{document}